\documentclass[aps, prd, reprint, showpacs, superscriptaddress, twocolumn, nofootinbib, amsmath]{revtex4-2}    

\usepackage{graphicx}
\usepackage{dcolumn}
\usepackage{bm}
\usepackage{natbib}
\usepackage{amsmath}
\usepackage{siunitx}
\usepackage [autostyle, english = american]{csquotes}
\MakeOuterQuote{"}
\usepackage{lipsum}
\usepackage{babel}
\usepackage{appendix}
\usepackage{xcolor}
\usepackage{soul}

\begin{document}
\title{Exploring resonantly produced mixed sterile neutrino dark matter models}
\author{Emma L. Horner}
\email{ehorner@sandiego.edu}
\affiliation{Department of Physics and Biophysics, University of San Diego, San Diego, CA 92110}

\author{Francisco {Mungia Wulftange}}
\affiliation{Department of Physics and Biophysics, University of San Diego, San Diego, CA 92110}

\author{Isabella A.\ Ianora}
\affiliation{Department of Physics and Biophysics, University of San Diego, San Diego, CA 92110}

\author{Chad T.\ Kishimoto}
\affiliation{Department of Physics and Biophysics, University of San Diego, San Diego, CA 92110}
\affiliation{Center for Astrophysics and Space Sciences, University of California, San Diego, La Jolla, CA 92093}

\date{\today}

\begin{abstract}

An unidentified 3.55 keV X-ray line in stacked spectra of galaxies and clusters raises the interesting possibility that it originates from the decay of sterile neutrino dark matter. In this work, we explore mixed sterile neutrino dark matter models that combine cold dark matter and warmer sterile neutrino dark matter produced through lepton number-driven active-to-sterile neutrino transformation. We analyze the sensitivity of the sterile neutrino spectra on active-sterile mixing and on initial neutrino lepton numbers. Furthermore, we assess the viability of these models with estimates of the number of subhalos formed as the host sites of satellite galaxies.

\end{abstract}

\maketitle

\section{Introduction}

The nature of dark matter, which comprises nearly all of the non-relativistic matter in the universe, remains a mystery. An interesting piece of the dark matter puzzle is the discovery of an unidentified $3.55~{\rm keV}$ X-ray line in stacked spectra of galaxies and clusters \cite{bulbul14, boy14}. One possible explanation for this X-ray line is the decay of sterile neutrino dark matter with a mass of $7.1~{\rm keV}$ that decays into the aforementioned X-ray photon and an equally energetic active neutrino. While there is vigorous debate surrounding the interpretation of this feature in the data, it remains an intriguing possibility that these X-ray spectra may probe particle physics beyond the Standard Model.

Sterile neutrinos are an attractive dark matter particle candidate \cite{Kusenko2009, steriledm, abazajian2017, boyarsky19}. Other than through gravitation, sterile neutrinos do not interact with Standard Model particles unless they mix with active neutrinos or are affected by Beyond Standard Model (BSM) interactions. They can be incorporated into a minimal extension of the Standard Model to account for non-zero neutrino masses \cite{Asaka2005,abs05}. There are a variety of proposed mechanisms to create the observed density of sterile neutrino dark matter through active-to-sterile neutrino scattering-induced transformation \cite{dw94, sf99a, afp} and other mechanisms with BSM interactions \cite{ak19, ap08, Petraki2008, kty10, st06, pfkk16}.

What does it mean if the X-ray line is the result of the decay of virialized sterile neutrino dark matter? The sterile neutrino mass is well determined as twice the X-ray line energy. The decay would imply active-sterile mixing, characterized by a vacuum mixing angle, $\sin^2 2 \theta$. Measurements of this mixing angle – {\it e.g.}, as in Refs.~\cite{bulbul14, boy14} – need to assume that all of the dark matter in the telescope’s field of view is sterile neutrino dark matter. While models can be constructed to create the observationally-inferred quantity of sterile neutrino dark matter, their resulting distributions and the subsequent structure formation that ensues may be at odds with observations on small scales.

The cold dark matter paradigm may have tension with observations on small scales \cite{bb17}, particularly related to dwarf galaxy counts and dark matter density profiles. Active-to-sterile neutrino transformation through the Dodelson-Widrow mechanism \cite{dw94} or the lepton-number driven Shi-Fuller mechanism \cite{sf99a} produce warmer dark matter distributions, which provides an attractive model to alleviate this tension. However, these distributions may be too warm to create the observed small scale structure \cite{ch17,sch15}, but this issue may be alleviated by other BSM mechanisms that produce colder spectra \cite{ak19, ap08, Petraki2008, kty10, st06, pfkk16}.

In this work, we consider mixed sterile neutrino dark matter models where the term ``mixed’’ plays a double meaning. On one hand, the 7.1 keV sterile neutrino quantum mechanically mixes with each of the three active neutrinos allowing for three distinct channels to resonantly produce sterile neutrino dark matter through the Shi-Fuller mechanism. On the other hand, we examine a statistical mixture of these warmer sterile neutrinos and cold dark matter to comprise the full observed dark matter content \cite{ak19, boyarsky09, baur17}. While there are multiple avenues to constrain these dark matter models with small scale structure \cite{zelko22}, we focus on the formation of dark matter subhalos that could be the site of Milky Way satellites, as in Refs.~\cite{ch17, sch15}. The goal of this work is to explore the phase space of possibilities of mixed sterile neutrino dark matter spectra produced by this model and to discuss constraints on these models through the lens of small scale structure formation. 

We are agnostic on the identity of the cold dark matter particle in these mixed sterile neutrino dark matter models, as long as their distribution can be treated as that of cold dark matter. The Shi-Fuller mechanism requires an initial neutrino lepton number that is many orders of magnitude larger than the baryon-to-photon ratio, yet small compared to any detectable limits from observations of the Cosmic Microwave Background (CMB) or Big Bang Nucleosynthesis (BBN) yields. There are many proposed mechanisms to create such a lepton number \cite{Laine2008, km22, mmr99, mcd00, ccg99}, and we are agnostic to which mechanism, so long as the asymmetries are created at $T \gtrsim {\rm GeV}$.

In Section \ref{sec:Calculations} we detail the production of sterile neutrino dark matter through lepton number-driven active-sterile neutrino transformation in the early universe and in Section \ref{sec:Results} we explore the possibilities and model sensitivity of the resulting spectra. In Section \ref{sec:Structure} we discuss constraining these mixed sterile neutrino dark matter models with small scale considerations, and draw some conclusions in Section \ref{sec:Conclusion}. We use natural units throughout, with $\hbar = c = k_B = 1$.

\section{Producing Sterile Neutrino Dark Matter} \label{sec:Calculations}

\subsection{Calculations}

In this section, we present the calculations related to the production of sterile neutrino dark matter from an active neutrino, $\nu_{\alpha}$, using the Shi-Fuller mechanism \cite{sf99a}. The sterile neutrino distribution function $f_{s}(\epsilon, x)$, evolves with a Boltzmann equation that accounts for the scattering-induced transformation between sterile neutrinos and an active neutrino species:

\begin{align}
    \label{eq:boltzman}
    \frac{\partial f_s}{\partial x}= \frac{1}{4}\Gamma_{\alpha}  \! & \sin^22\theta_{m}  \times \bigg(1 + \bigg(  \frac{\Gamma_{\alpha} l_m}{2}\bigg)^2\bigg)^{-1}  \nonumber \\ & \times (f_{\alpha}(\epsilon, x) - f_s(\epsilon, x)) \times {\frac{dt}{dx}},
\end{align}
where $x = 1/T$, the inverse of the plasma temperature, is our time-like independent variable. Comoving neutrino energy, $\epsilon$, is defined by $\epsilon = p_{\nu}/T_{\mathrm{cm}}$: the neutrino momentum scaled by the comoving temperature,  \mbox{$T_{\mathrm{cm}} \propto a^{-1}$}, where $a$ is the scale factor.  We write the sterile distribution functions in terms of the scaled energy because the scaled energy of a freely-streaming particle remains unchanged in an expanding universe, so the expansion does not directly affect the distribution function. $f_{\alpha}(\epsilon, x)$ is the active neutrino distribution function, which we treat as a Fermi-Dirac spectrum with the plasma temperature. $\Gamma_{\alpha}$ is the active neutrino scattering rate. For the scattering rate, we use the forms in Ref.~\cite{afp} which are consistent with scattering off thermal populations of all three active neutrino species and the thermally produced charged leptons.

The production of sterile neutrino dark matter occurs in the rapidly expanding early universe. In the homogeneous and isotropic early universe, the entropy in a comoving volume is conserved and the evolution of the scale factor is, 
\begin{equation}
    \label{eq:scale factor evolution}
    \frac{da}{dx}= \frac{a}{x} - \frac{a}{3g_{*s}(x)}\frac{dg_{*s}(x)}{dx},
\end{equation}
where $g_{*s}(x)$ is the effective number of entropic degrees of freedom. 

To relate rates to $x$, as in Eq.~\ref{eq:boltzman}, we use the Friedmann Equation,
\begin{equation}
    \label{eq:time temp relation}
    \frac{dt}{dx}= \frac{x^2m_\mathrm{pl}}{a\pi}\bigg(\frac{8\pi}{90}g_*(x)\bigg)^{-\frac{1}{2}}\frac{da}{dx},
\end{equation}
where $m_{\mathrm{pl}} = 1.22 \times 10^{22}$~MeV is the Planck mass, and $g_{*}(x)$ is the effective number of relativistic degrees of freedom. (We use the results from Ref.~\cite{ls06} for $g_* (T)$ and $g_{*s} (T)$ in our calculations.)

 The production of sterile neutrinos described in Eq.~\ref{eq:boltzman} uses the Quantum Zeno approximation, that connects coherent in-medium oscillations with de-coherent scattering-induced sterile neutrino production \cite{afp, fv97}. These oscillations are affected by forward-scattering interactions with the other particles in the plasma and can be characterized by the oscillation length and effective matter mixing angle.
 The oscillation length,

\begin{equation}
    \label{eq:oscillaiton length}
    l_m=\Big(\Delta^2\sin^22\theta + [\Delta\cos2\theta - V^D - V^T]^2\Big)^{-\frac{1}{2}},
\end{equation}
is the physical distance traveled by a neutrino as it completes an oscillation. The effective matter mixing angle describes how neutrinos oscillate in the hot and dense primordial plasma, taking into account the neutrino interactions. It is defined by,
\begin{equation}
    \label{eq:matter mixing angle}
    \sin^22\theta_m= \frac{\Delta^2\sin^22\theta}{\Delta^2\sin^22\theta + [\Delta\cos2\theta - V^D - V^T]^2}, 
\end{equation}
where $\sin^22\theta$ is the vacuum mixing angle, \mbox{$\Delta = m_s^2/2\epsilon T_{\mathrm{cm}}$}, and $m_s$ is the sterile neutrino mass.

The forward-scattering potential is accounted for by introducing two potentials, $V^D$ and $V^T$. The first of these is the density potential:

\begin{equation}
    \label{eq:denity potential}
    V^D= \frac{2\sqrt{2}\zeta(3)}{\pi^2}G_F T^3 \mathcal{L}^{\alpha},
\end{equation}
where $\zeta(3)=1.2020569$ and $G_F = 1.166 \times 10^{-11}~ \mathrm{MeV}^{-2}$ is the Fermi constant. The lepton potential is \mbox{$\mathcal{L}^{\alpha} = 2L_{\nu_{\alpha}} + \Sigma_{\beta} L_{\nu_{\beta}}$} with $\alpha \neq \beta$ where $L_{\nu}$ is the neutrino lepton number. $V^D$ describes the effect of a non-zero neutrino number density on neutrino-neutrino interactions. The density potential depends on asymmetries between the neutrinos and anti-neutrinos.

The second potential that arises from neutrino interactions, $V^T$, is the thermal  potential. It is described by, 
\begin{align}
    \label{eq:thermal potential}
    V^T  =- &\frac{8\sqrt{2} G_F  \epsilon T_{\mathrm{cm}}}{3m_Z^2}(\rho_{\nu_{\alpha}}+\rho_{\bar\nu_{\alpha}}) \nonumber \\
    & -\frac{8\sqrt{2} G_F \epsilon T_{\mathrm{cm}}}{3m_W^2}(\rho_{\alpha}+\rho_{\bar\alpha}), 
\end{align}
where $\rho_{\nu_{\alpha}}$ and $\rho_{\bar\nu_{\alpha}}$ are the neutrino and anti-neutrino energy densities for flavor $\alpha$. $\rho_{\alpha}$ and $\rho_{\bar\alpha}$ are the energy densities for the charged lepton and antilepton that correspond to flavor $\alpha$. $V^T$ describes the influence of thermal distributions of particles on the forward scattering of neutrinos \cite{nr88}. 

We also follow the transformation of anti-neutrinos $\bar\nu_{\alpha}$ into $\bar\nu_{s}$. The $\bar\nu_{s}$ distribution function evolves with a Bolztmann equation similar to Eq.~\ref{eq:boltzman}. For anti-neutrino transformation, we use the anti-neutrino distribution functions $\bar f_{s}(\epsilon, x)$ and $\bar f_{\alpha}(\epsilon, x)$, and the anti-neutrino forward scattering potentials are $\overline V^D = - V^D$ and $\overline V^T = V^T$. This is because $V^D$ is anti-symmetric when swapping particles and anti-particles , while $V^T$ is symmetric.

As active neutrinos transform into steriles, their corresponding lepton numbers decrease. The lepton number evolution is described by
\begin{align}
    \label{eq:lep evo}
    \frac{dL_{\nu_{\alpha}
    }}{dx}=  -\frac{1}{n_{\gamma}} \frac{T_{\mathrm{cm}}^3}{2\pi^2}\bigg(\int_0^{\infty} \! & \epsilon^2 \frac{df_{s}}{dx}d\epsilon -  \int_0^{\infty} \! \epsilon^2 \frac{d\bar f_{s}}{dx}d\epsilon \bigg) \nonumber \\ &  - L_{\nu_{\alpha}} \bigg(\frac{3}{a}\frac{da}{dx}+\frac{3}{T}\frac{dT}{dx}\bigg).
\end{align}
where $n_{\gamma}$ is the number density of photons.
The integral terms describe how the production of sterile neutrinos and anti-neutrinos change the lepton number as actives transform into steriles. The last term takes into account the effect of dilution. All three lepton numbers evolve due to dilution, but only flavors that transform into steriles are affected by the production terms.

\subsection{Mixed Sterile Neutrino Dark Matter Models}

We consider two separate scenarios for the production of sterile neutrino dark matter: the ``one-to-one model'' in which one active neutrino flavor transforms into a distribution of sterile neutrinos according to Eq.~\ref{eq:boltzman}, and the ``three-to-one model'' in which all three active neutrino flavors transform into a sterile distribution simultaneously with non-zero mixing angles in all three active to sterile channels. In the three-to-one model, $\partial f_s / \partial x$ has three terms, each as shown in Eq.~\ref{eq:boltzman}, but specific to each active-to-sterile production channel. Each active neutrino lepton number evolves independently according to Eq.~\ref{eq:lep evo} with contributions from the specific active-to-sterile production channel and dilution.

The one-to-one model is described by a single active-to-sterile mixing angle and three initial neutrino lepton numbers. The three-to-one model is described by three mixing angles, one for each active-to-sterile channel and three initial lepton numbers. Each mixed sterile neutrino dark matter model is supplemented by the appropriate density of cold dark matter to match the Planck dark matter measurements \cite{Planck2018}.

\begin{figure}[t!]
    \centering
   \includegraphics[width=7cm]{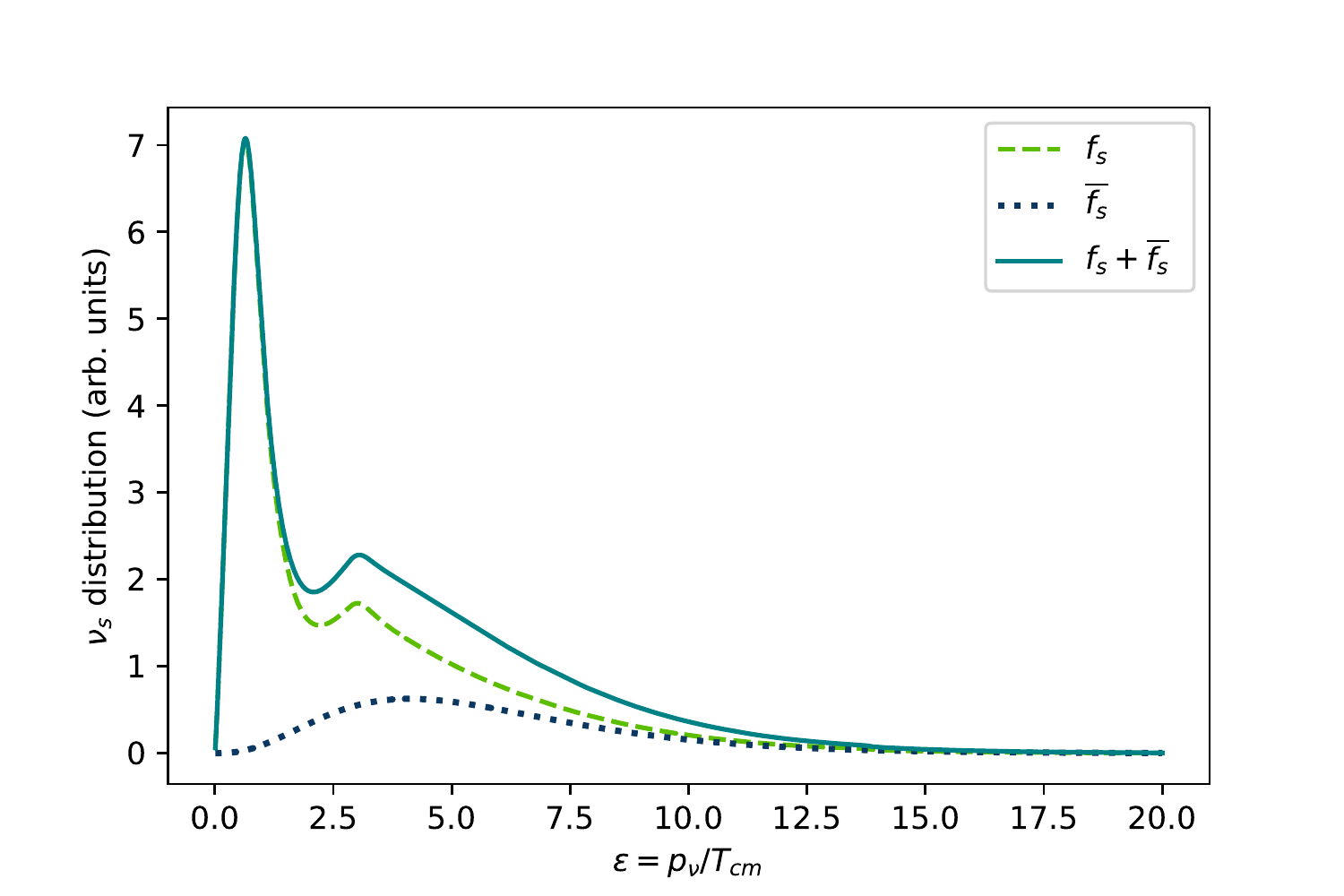}
    \caption{Sterile neutrino dark matter spectrum as a function of comoving neutrino energy from the one-to-one model. The solid curve represents total dark matter produced, which is the sum of the sterile neutrino production (dashed curve) and the anti-neutrino production (dotted curve). Here, production is only in the muon to sterile sterile channel with $\sin^22\theta = 5.3 \times 10^{-10}$ and initial lepton number in all three flavors, $L_{\nu_e0} = L_{\nu_{\mu}0} = L_{\nu_{\tau}0} = 12 \times 10^{-4}$.}
    \label{fig: spectra-justmu}
\end{figure}

\section{Results} \label{sec:Results}
In this section we will discuss various scenarios for lepton number-driven mixed sterile neutrino dark matter in the early universe using the Shi-Fuller mechanism. For each model, we explore the sterile neutrino distributions and the evolution of the lepton numbers. In particular we are interested in how the initial lepton numbers and mixing angles affect the sterile neutrino dark matter distributions.

\subsection{One-to-one Model}
\begin{figure}[t!]
    \centering
   \includegraphics[width=7cm]{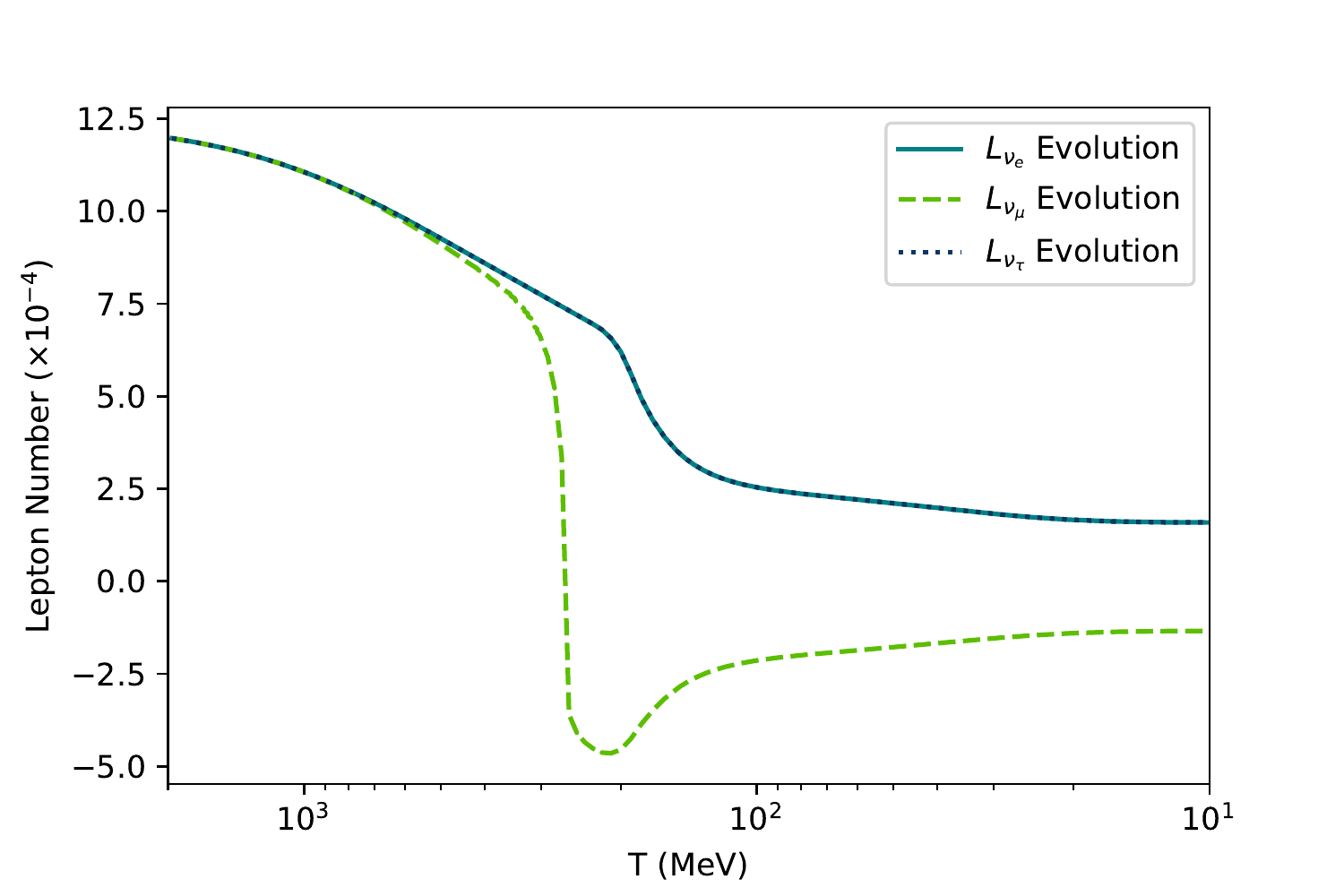}
    \caption{Lepton number evolution as a function of decreasing universe temperature corresponding to the model in Fig.~\ref{fig: spectra-justmu}. The solid curve shows the electron neutrino lepton number evolution, the dashed curve shows the muon neutrino lepton number evolution, and the dotted curve shows the tau  neutrino lepton number evolution. The electron and tau neutrino lepton number evolution are identical. }
    \label{fig: lep-justmu}
\end{figure}

As an illustration, we show the results of a one-to-one model where a distribution of sterile neutrino dark matter is produced solely through the muon to sterile channel. Fig.~\ref{fig: spectra-justmu} shows the sterile neutrino dark matter spectrum and Fig.~\ref{fig: lep-justmu} shows the evolution of the active neutrino lepton numbers.

The lepton number-driven resonant production can be seen by the resonant peaks in the colder part of the spectrum in Fig.~\ref{fig: spectra-justmu}, where the neutrinos have lower energies \cite{kf08}.
The production of the resonant peaks corresponds to the sharp decline in the muon neutrino lepton number as the muon neutrinos rapidly transform into sterile neutrinos, which can be seen in Fig.~\ref{fig: lep-justmu} at $ T \approx  300~\mathrm{MeV}$. The muon neutrino lepton number is the only one out of the three active flavors that has this sharp decline because this model only has production in the muon to sterile channel.

When the lepton number has been depleted there is no more resonant production, but non-resonant production continues to transform active neutrinos into sterile neutrinos. This non-resonant production can be seen best in the anti-neutrino distribution because the anti-neutrinos do not have the resonant condition so all of their production is non-resonant production. A similar distribution from non-resonant production occurs in the neutrino distribution, but it is obscured by resonant production except for an exponential tail at high scaled energy.

Each active neutrino lepton number initially declines in sync until  $ T \approx 500~\mathrm{MeV}$, experiencing the effect of dilution. 
This effect manifests as the universe temperature decreases, the energy required for particle-anti-particle creation becomes higher than the energy available in the plasma, and the particle creation rate is exponentially suppressed. Consequently, the thermodynamic equilibrium between creation and annihilation favors the annihilation of more massive quark degrees of freedom as their rest mass energy is thermalized into the plasma. The denominator of the lepton number is the number density of photons, 
\begin{equation}
    \label{eq:lepton number}
    L_{\nu_{\alpha}}= \frac{n_{\nu_{\alpha}} - n_{\bar\nu_{\alpha}}}{n_{\gamma}},
\end{equation}
 so overall the lepton numbers asymptote to zero because this effect causes the number density of photons to increase, but does not change the neutrino-anti-neutrino asymmetry. The effects of dilution can be discerned at $ T = 180~\mathrm{MeV}$ where the rapid loss of asymptotically free quark degrees of freedom at the QCD transition causes the lepton numbers to sharply trend towards zero.

\begin{figure}[t!]
    \centering
    \includegraphics[width=7cm]{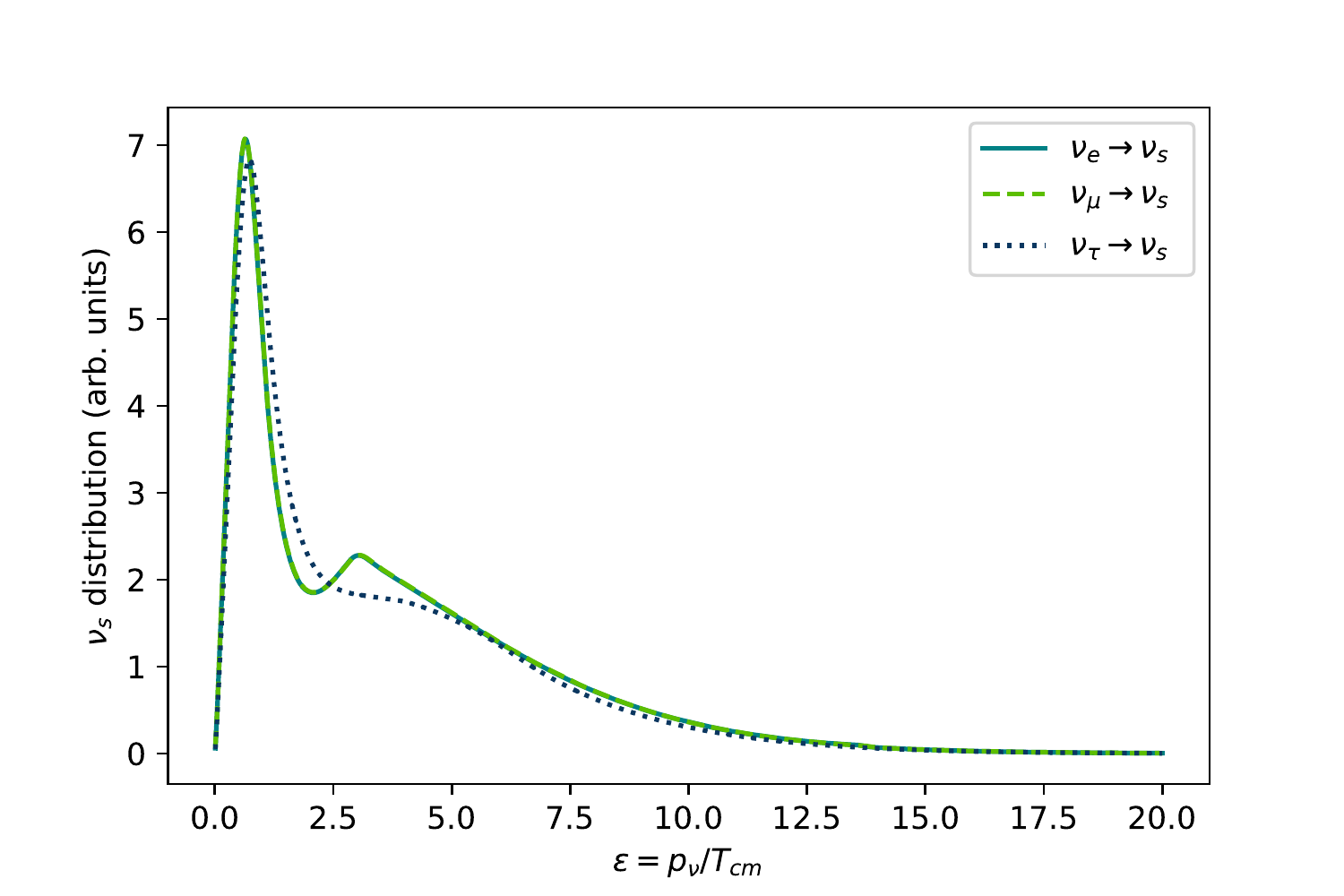}
    \caption{Sterile neutrino dark matter spectrum as a function of comoving neutrino energy for the three one-to-one models. Electron to sterile transformation is shown by the solid line, muon to sterile is shown by the dashed line, and the dotted line shows the tau to sterile transformation. In this calculation we use the same values for the mixing angle and initial lepton numbers as in Fig.~\ref{fig: spectra-justmu}. }
    \label{fig: spectra 1 to 1}
\end{figure}

\begin{figure}[t!]
    \centering
    \includegraphics[width=7cm]{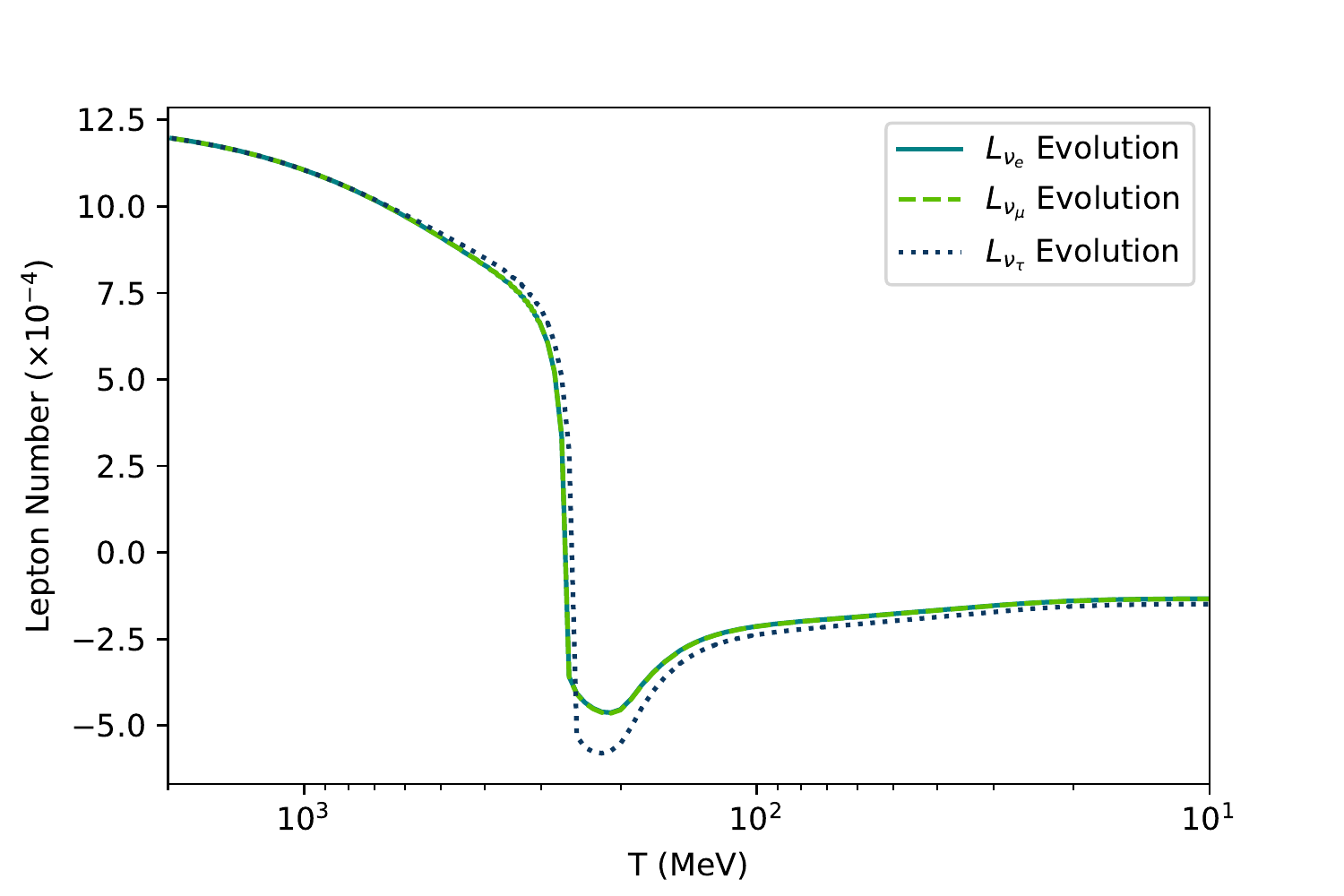}
    \caption{Lepton number evolution as a function of decreasing universe temperature corresponding to the model in Fig.~\ref{fig: spectra 1 to 1}. The dashed curve is the evolution of $L_{\nu_{\mu}}$ in the muon to sterile production channel, as in Fig.~\ref{fig: lep-justmu}. For this channel, the $L_{\nu_e}$ and $L_{\nu_{\tau}}$ are not shown on this figure, but they evolve as shown in Fig.~\ref{fig: lep-justmu}. The solid curve is the evolution of $L_{\nu_e}$ in a one-to-one electron to sterile channel. The $L_{\nu_{\mu}}$ and $L_{\nu_{\tau}}$ evolution in this channel are not shown in this figure, but they evolve as the non-transforming lepton numbers in Fig.~\ref{fig: lep-justmu} ({\it e.g.,} $L_{\nu_e}$ and $L_{\nu_{\tau}}$). Likewise for the tau to sterile channel (dotted curve). }
    \label{fig: lep evo 1 to 1}
\end{figure}

There are three distinct one-to-one models: the electron, the muon, and the tau to sterile production models. Fig.~\ref{fig: spectra 1 to 1} and Fig.~\ref{fig: lep evo 1 to 1} show the sterile neutrino dark matter spectra and the lepton numbers evolution of these models, respectively, using the same mixing angle and initial lepton numbers as in Fig.~\ref{fig: spectra-justmu} and Fig.~\ref{fig: lep-justmu}. The results for the electron (solid curve) and muon (dashed curve) to sterile channels are nearly identical and produce 22.6\% of the total dark matter, while the tau (dotted curve) to sterile production produces  22.1\%. 

The similarities and differences between these models occur because we see in Fig.~\ref{fig: lep evo 1 to 1} that most of the resonant production occurs around $ T \approx 300 ~\mathrm{MeV}$. Here, there are large thermal populations of $e^{\pm}$ and $\mu^{\pm}$ but not $\tau^{\pm}$. As a result, the scattering rate in the electron and the muon production channels will be larger than in the tau channel. A larger scattering rate supresses resonant production due to the Quantum Zeno effect and enhances non-resonant production. In addition, the thermal potential terms are different: \mbox{$|V^T_{\tau}| < |V^T_e| = |V^T_{\mu}|$}. The result is differences in how resonances sweep through the active neutrino distribution. This leads to resonant production occurring later in the tau channel compared to the electron and muon channels, at lower temperatures. This later production enhances resonant production with reduced Quantum Zeno suppression and a slower expansion rate.

\subsection{Three-to-one Model}

\begin{figure}[t!]
    \centering
    \includegraphics[width=7cm]{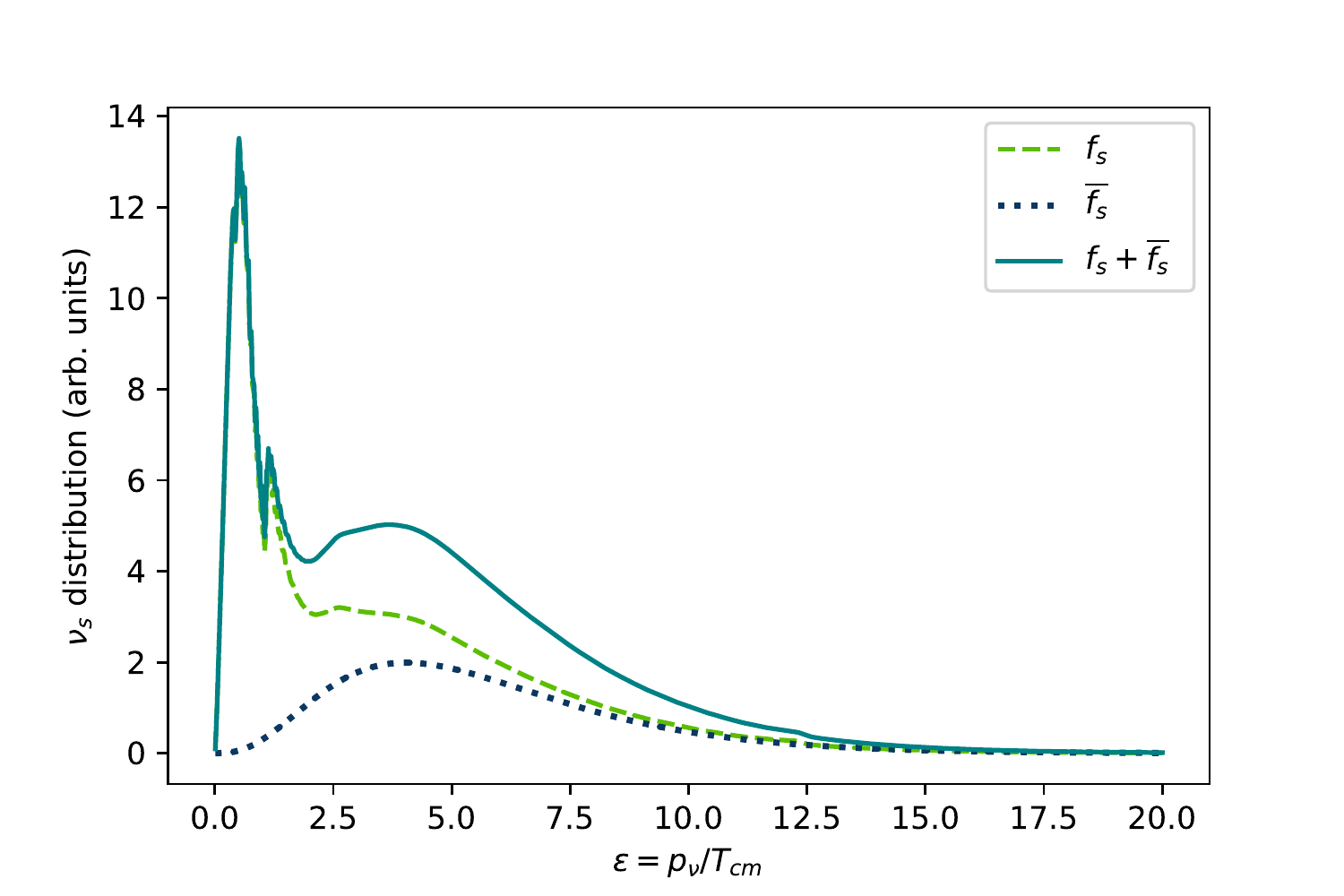}
    \caption{Sterile neutrino dark matter spectrum for the three-to-one model. The dashed line is the dark matter produced by sterile neutrinos, the dotted line is that produced by anti-neutrinos, and the solid line is the sum of the two. In this calculation there are equal mixing angles in all three active-to-sterile channels and equal initial lepton numbers. For the sake of comparison, we use the same values as in Fig.~\ref{fig: spectra-justmu}. }
    \label{fig: spectra 3 to 1}
\end{figure}

\begin{figure}[t!]
    \centering
    \includegraphics[width=7cm]{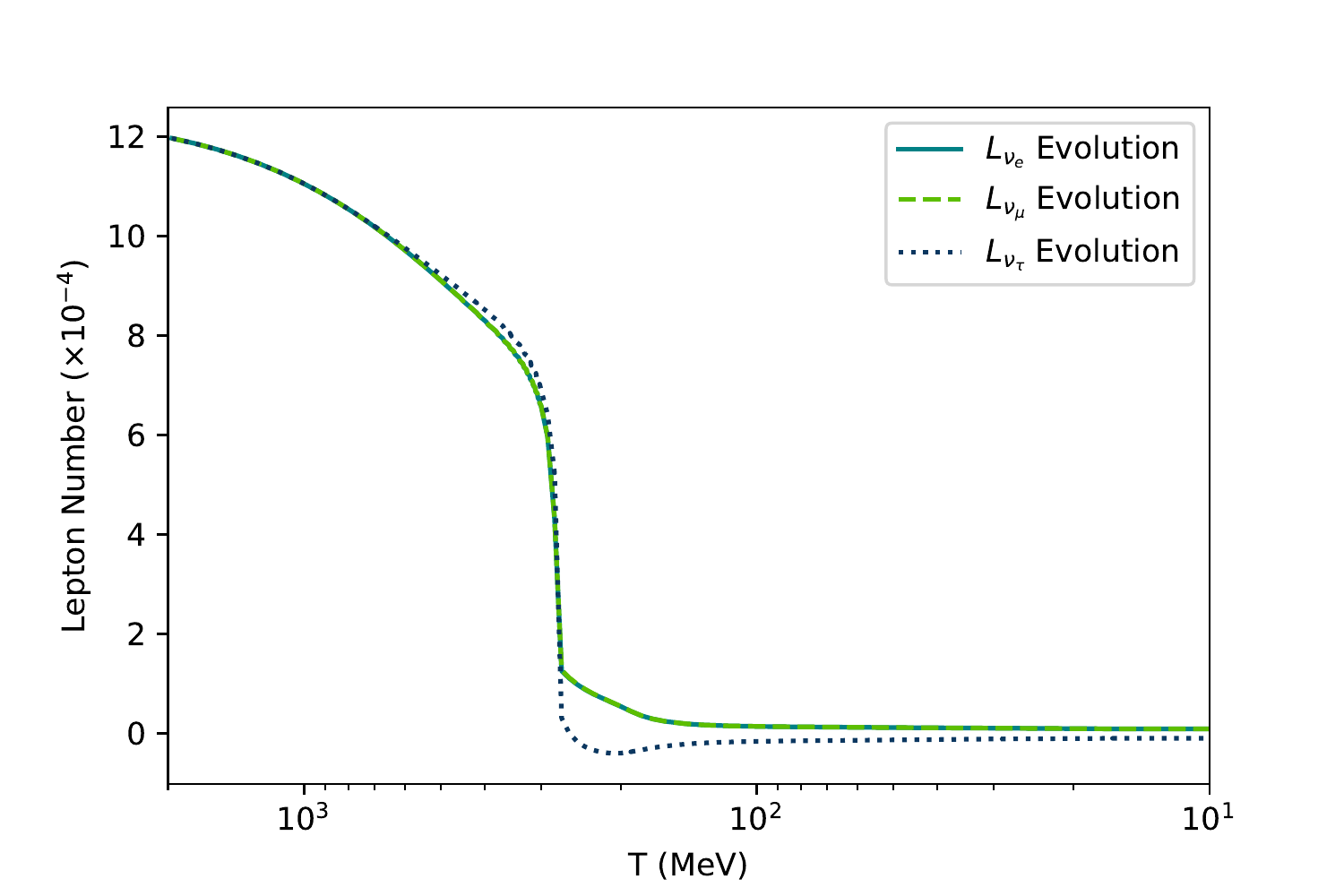}
    \caption{Neutrino lepton number evolution for the three-to-one model shown in Fig.~\ref{fig: spectra 3 to 1}. The solid curve shows the evolution of the electron neutrino lepton number, the dashed curve shows the evolution of the muon neutrino lepton number, and the dotted curve shows the tau neutrino lepton number evolution.}
    \label{fig: lep evo 3 to 1}
\end{figure}

We now compare and contrast the three distinct one-to-one models with a three-to-one model where all three active flavors simultaneously transform into sterile neutrinos. As before, Fig.~\ref{fig: spectra 3 to 1} and Fig.~\ref{fig: lep evo 3 to 1} show the sterile neutrino dark matter spectra and lepton numbers evolution, respectively, of a three-to-one model. This model uses the same mixing angles and initial lepton numbers as shown in the one-to-one models above, and produces 51.4\% of the total dark matter. While there are similarities between the three-to-one model and a superposition of the three one-to-one models in Fig.~\ref{fig: spectra 1 to 1} and Fig.~\ref{fig: lep evo 1 to 1}, their primary differences are the result of the fact that all three lepton numbers are depleted by resonant transformation. The density potential, $V^D$ (Eq.~\ref{eq:denity potential}), is proportional to the lepton potential, which incorporates all three neutrino lepton numbers. As a result, the resonant production in all three channels are linked, as can be seen in Fig.~\ref{fig: lep evo 3 to 1} with the nearly simultaneous sharp decline in all three lepton numbers. The result is less resonant production for each individual channel as the lepton potential depletes faster in this three-to-one model. 

\begin{figure*}
    \centering
    \includegraphics[width=0.9\textwidth]{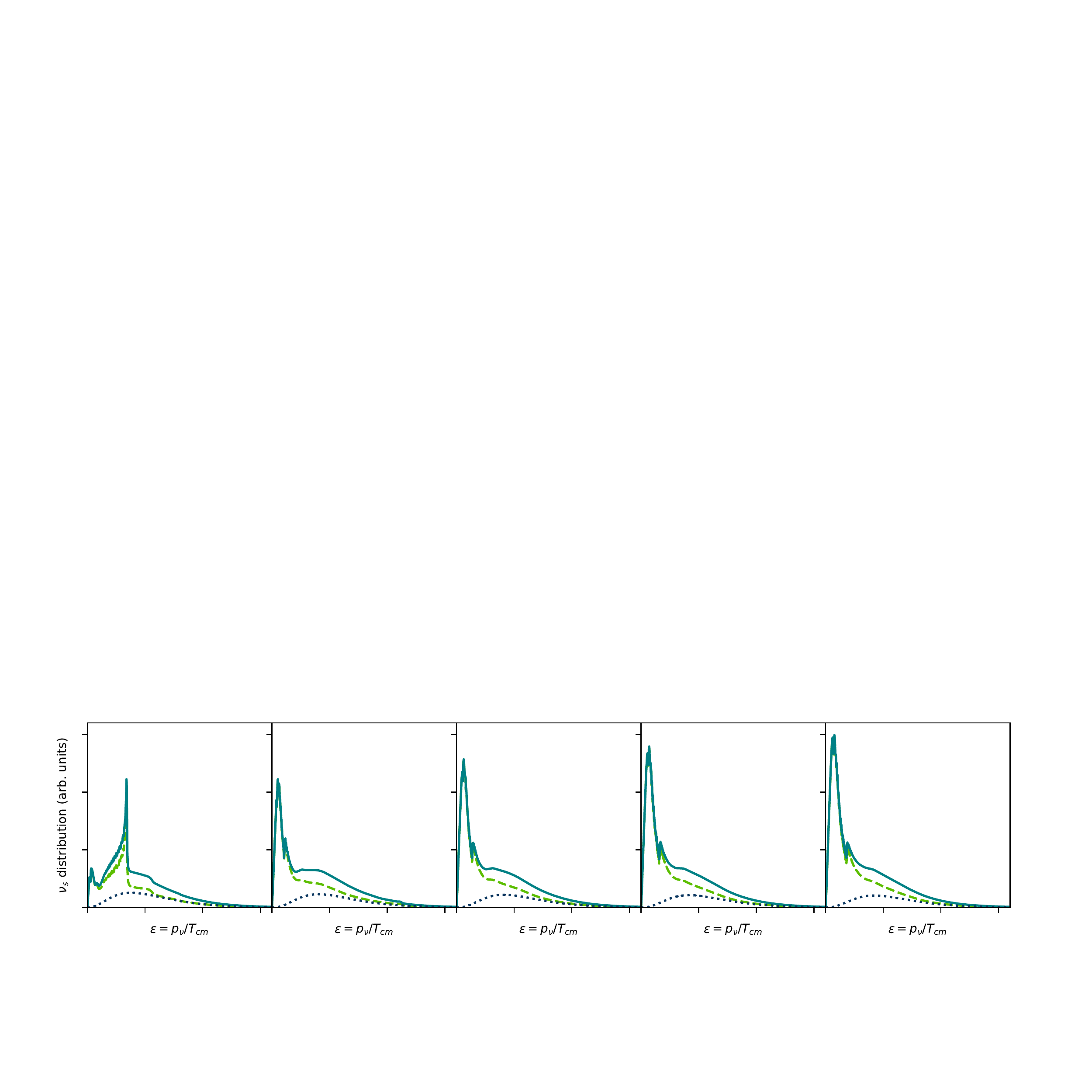}
    \caption{Sample of five sterile neutrino dark matter spectra with equal mixing angles for all three flavors, $\sin^22\theta = 10^{-9.5}$. The initial lepton numbers increase linearly from left to right with $L_{\nu_e0} = L_{\nu_{\mu}0} = L_{\nu_{\tau}0} = 7.6 \times 10^{-4}, 11 \times 10^{-4}, 13 \times 10^{-4}, 15 \times 10^{-4}$, and $17 \times 10^{-4}$. As in Fig.~\ref{fig: spectra 3 to 1}, the total spectrum (solid curve) is the sum of neutrino (dashed) and anti-neutrino (dotted) production.}
    \label{fig:rowplotlep}
\end{figure*}

The three-to-one models are fully described by their mixing angles and initial lepton numbers, with each set of parameters producing a unique spectrum and mix of sterile neutrino dark matter and cold dark matter. To explore the effects of varying the initial lepton number, Fig.~\ref{fig:rowplotlep} shows a sample of five initial lepton numbers and a constant mixing angle. We find that as the initial lepton number increases, there is more resonant production and the resonances shift to slightly higher $\epsilon$ values. Larger initial lepton numbers allow for more active-sterile transformation before the lepton number is depleted, ceasing resonant production. In addition, the anti-neutrino production in Fig.~\ref{fig:rowplotlep} shows that the non-resonant production is largely unaffected by changing the initial lepton number.

\begin{figure*}
    \centering
    \includegraphics[width=0.9\textwidth]{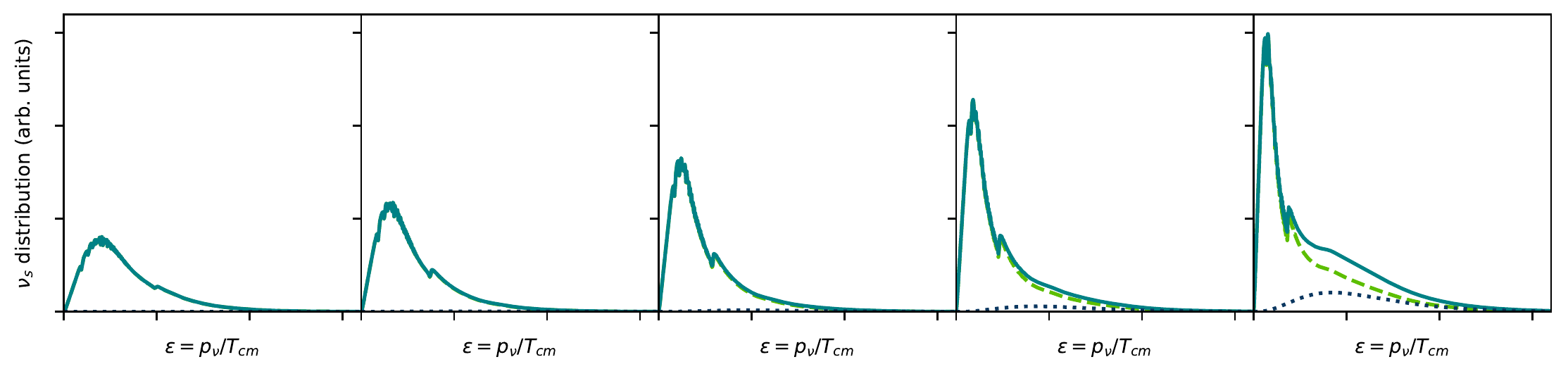}
    \caption{Sample of five sterile neutrino dark matter spectra with equal initial lepton numbers, $L_{\nu_e0} = L_{\nu_{\mu}0} = L_{\nu_{\tau}0} = 17 \times 10^{-4}$. From left to right, the mixing angles (equal for all three flavors) increase logarithmically with $\sin^22\theta = 10^{-11.5}, 10^{-11}, 10^{-10.5}, 10^{-10}, 10^{-9.5}$. As in Fig.~\ref{fig: spectra 3 to 1}, the total spectrum (solid curve) is the sum of neutrino (dashed) and anti-neutrino (dotted) production.}
    \label{fig:rowplotmix}
\end{figure*}

To explore the effects of varying the mixing angle, Fig.~\ref{fig:rowplotmix} shows a sample of five mixing angles and a constant initial lepton number. When we increase the mixing angle there is an increase in both resonant and non-resonant production. The boost in resonant production can be seen in the low-$\epsilon$ side of the spectrum and is the result of broadening of the resonance widths. The increase in non-resonant production is easily distinguished from the anti-neutrino distribution. The non-resonant production rate is proportional to the mixing angle.

\section{Formation of Small-Scale Structure} \label{sec:Structure}

One method to constrain our mixed sterile neutrino dark matter models is to consider their effects on the formation of small scale structure.  Purely cold dark matter may overproduce the dark matter subhalos that are the site of satellite galaxies \cite{horiuchi16}, while purely resonantly produced sterile neutrino dark matter underproduces these subhalos \cite{sch16, ch17}. For our mixed sterile neutrino dark matter models, we will use the number of subhalos produced as a simple constraint on the viability of these models.

We use the CLASS code \cite{class112} to self-consistently calculate the matter power spectrum from the mixed sterile neutrino dark matter model, using the $\Lambda \mathrm{CDM}$ model values from Planck \cite{Planck2018}. We treat the sterile neutrinos produced as non-cold dark matter in CLASS with the $\Omega_s h^2$ and spectrum produced as described in the previous section. The remainder of the observed dark matter, we treat as cold dark matter in CLASS.

\begin{figure}[t!]
    \centering
   \includegraphics[width=7cm]{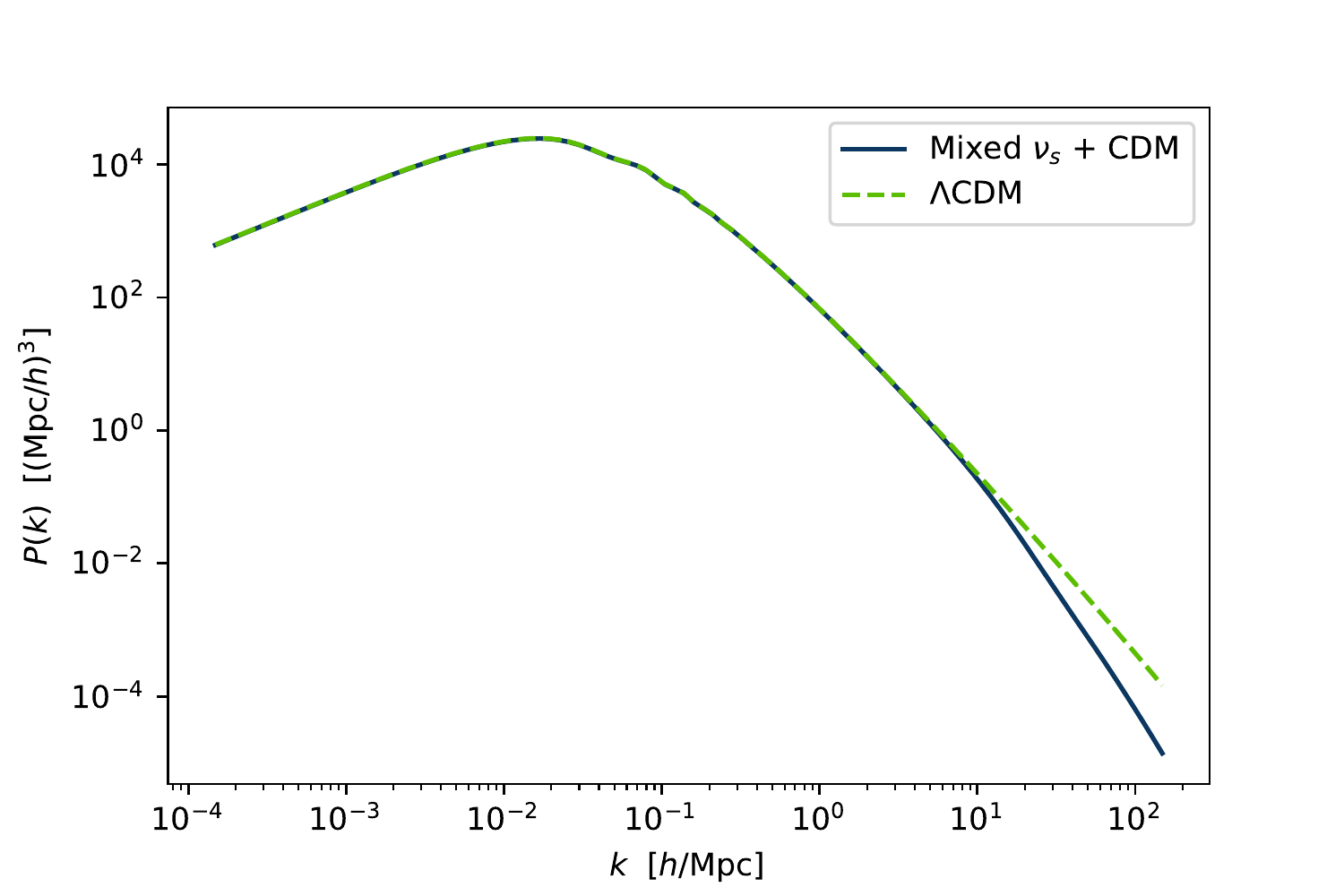}
    \caption{Comparison of matter power spectra for mixed sterile neutrino dark matter model in Fig.~\ref{fig: spectra 3 to 1} (solid curve) and $\Lambda \mathrm{CDM}$ (dashed curve). }
   \label{fig:matter power spectrum}
\end{figure}

Fig.~\ref{fig:matter power spectrum} compares the matter power spectrum for $\Lambda\mathrm{CDM}$ (dashed curve) with the mixed sterile neutrino dark matter model discussed in the previous section. Both matter power spectra are calculated with CLASS using the same cosmological parameters, except the mixed sterile neutrino dark matter model is comprised of 48.6\% cold dark matter and 51.4\% non-cold dark matter with the spectrum shown in Fig.~\ref{fig: spectra 3 to 1}. Importantly, the warm tail of the sterile neutrino dark matter distribution, primarily created by non-resonant production, suppresses the high-$k$ tail of the matter power spectrum.  This reduces the number of subhalos produced in such a model.

We used the methods described in Ref.~\cite{sch15} to estimate the number of subhalos with mass greater than $10^8 h^{-1} M_\odot$ that are produced by the mixed sterile neutrino dark matter models. While this analysis is an oversimplification --- the linear calculation of the matter power spectrum cannot capture the non-linear complexity of the formation of these subhalos --- its simplicity allows for a first constraint on the dark matter distributions. 

Ref.~\cite{sch15} suggests a figure of merit of sixty subhalos. Models that produce fewer than 60 subhalos cannot reproduce the expected number of satellite galaxies of the Milky Way as each satellite galaxy ought have its own subhalo. A viable mixed sterile neutrino dark matter model needs to produce at least 60 subhalos, but this analysis does not consider whether satellite galaxies form in each subhalos nor whether feedback effects reduce the number of subhalos. The model shown in Fig.~\ref{fig: spectra 3 to 1} 
produced 80 subhalos and would thus be a viable model.

\section{Discussion and Conclusions} \label{sec:Conclusion}

\begin{figure}[t!]
    \centering
    \includegraphics[width=9cm]{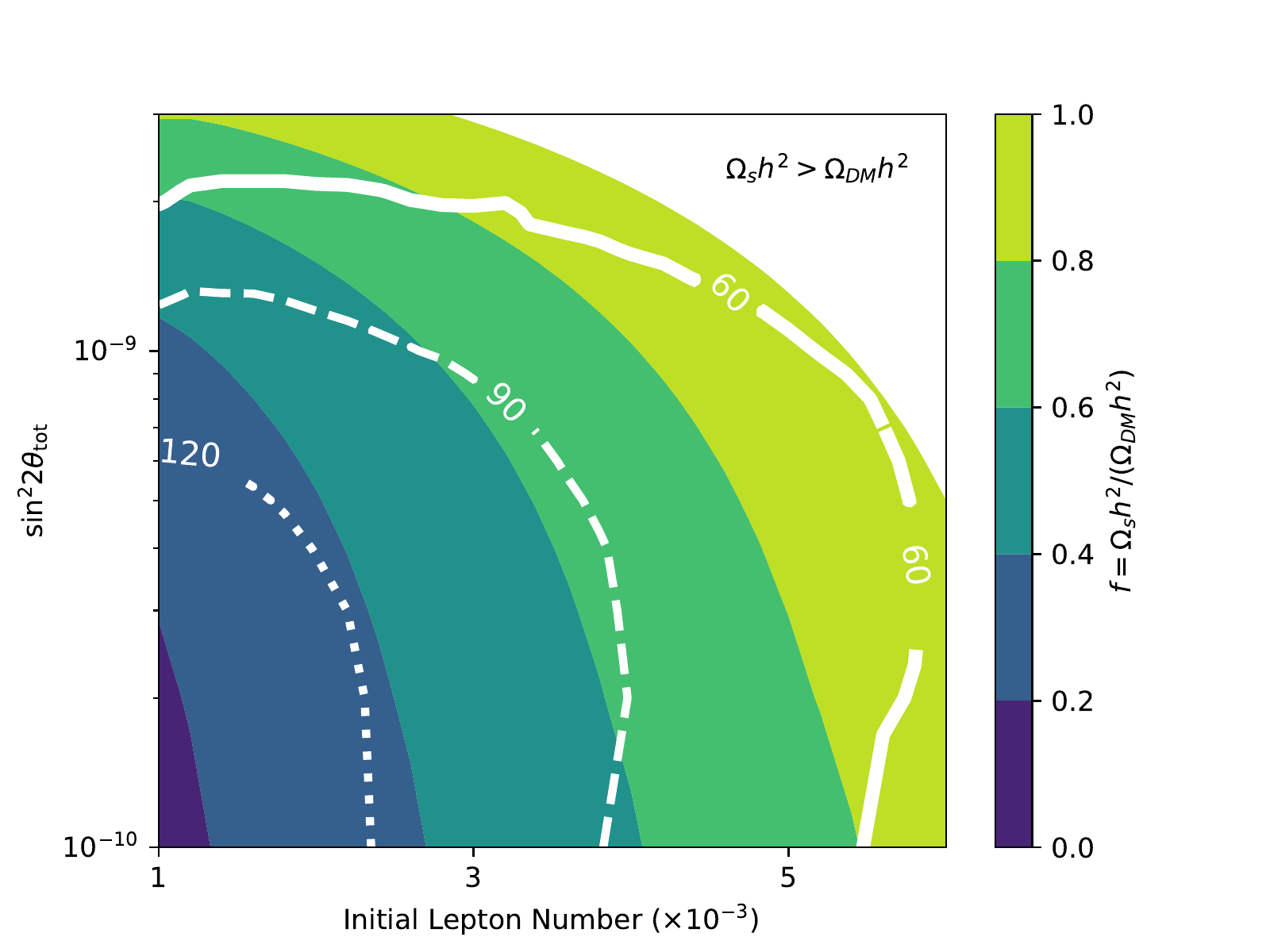}
    \caption{Results from three-to-one models with three equal mixing angles and equal initial lepton numbers. These models are parametrized by the total mixing angle, $\sin^22\theta_{\mathrm{tot}}$, and the initial lepton number in one flavor. The color contours represent the fraction of the dark matter that is sterile neutrino dark matter. The white contours show the number of subhalos produced by the models. Models that overproduce sterile neutrino dark matter, $\Omega_sh^2 > \Omega_{\mathrm{DM}}h^2$, or underproduce subhalos (less than 60) can be excluded.  }
    \label{fig: contour plot suhalos}
\end{figure}

\begin{figure}[t!]
    \centering
    \includegraphics[width=9cm]{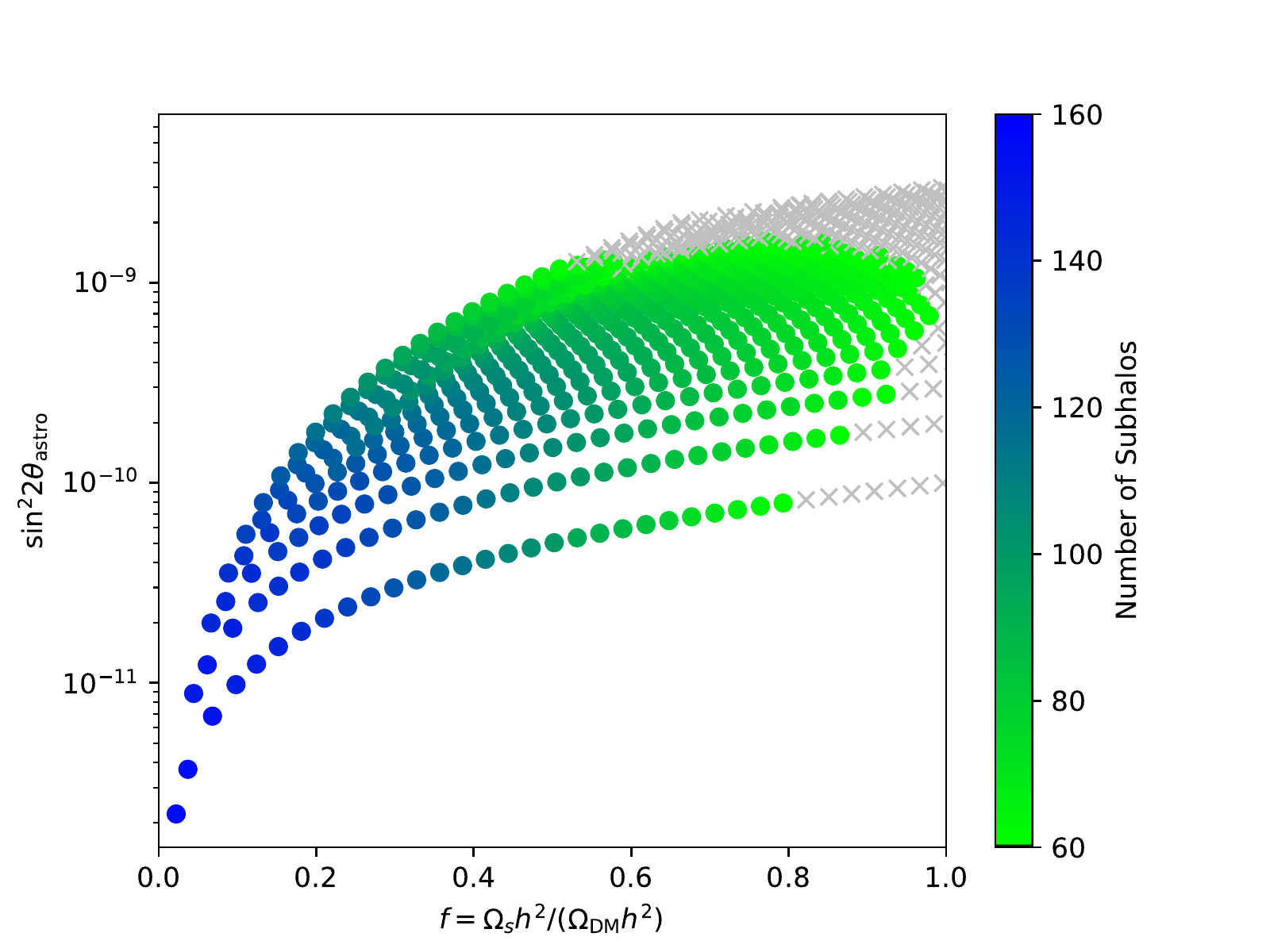}
    \caption{Another perspective on the models shown in Fig.~\ref{fig: contour plot suhalos}. Each three-to-one model is shown with its astrophysically inferred mixing angle, $\sin^22\theta_{\mathrm{astro}}$, fraction of sterile neutrino dark matter, $f$, and the number of subhalos. The color scheme shows the number of subhalos produced: ranging from 160, consistent with 100\% cold dark matter ($f=0$), to 60, which is our minimum constraint.}
    \label{fig: scatter plot mixang}
\end{figure}

 Fig.~\ref{fig: contour plot suhalos} and Fig.~\ref{fig: scatter plot mixang} summarize the results of the three-to-one model with equal initial lepton numbers and mixing angles in all three active-to-sterile channels. These models are characterized by two fundamental parameters. One is the total mixing angle,

\begin{equation}
    \label{eq:total mixang}
    \sin^22\theta_\mathrm{tot} = \sin^22\theta_e + \sin^22\theta_{\mu} + \sin^22\theta_{\tau},
\end{equation}
which is the effective active-sterile mixing angle. The other parameter is the initial lepton number. The total mixing angle connects to neutrino mass and mixing models, but the values explored here are far too small to be probed in terrestrial experiments, although they have many interesting consequences in core collapse supernovae and pulsar kicks \cite{Fuller2003, kusenko06, Kusenko1999}. The initial lepton number connects to cosmological models that create these asymmetries, but the resulting lepton numbers are far too small to be probed through their influences on BBN yields and the CMB \cite{gfkp17, lrv05, kssw01, shimon10}.

Fig.~\ref{fig: contour plot suhalos} explores this mixed dark matter model parameter space. The shaded contours correspond to the fraction of the total dark matter that is sterile neutrino dark matter, $f = \Omega_sh^2 / \Omega_{\mathrm{DM}}h^2$, where $\Omega_sh^2$ is the sterile neutrino dark matter density produced in these models and $\Omega_{\mathrm{DM}}h^2$ is the Planck-measured dark matter density. The white contours show the number of subhalos in a Milky Way-sized host galaxy produced in the mixed sterile neutrino dark matter models. 

First, we exclude the parameter space that overproduces sterile neutrino dark matter, $\Omega_sh^2 > \Omega_{\mathrm{DM}}h^2$. These models are inconsistent with current dark matter measurements. Of the models that remain, viable models lie below the 60 subhalo contour line, indicating that the model produced at least 60 sites for satellite galaxies. As discussed above, the 60 subhalo contour should be approached with reasonable circumspection. Rather than as a hard line of demarcation, this region-- which in Fig.~\ref{fig: contour plot suhalos} includes models with 60-100\% sterile neutrino dark matter-- should be thought of as models of interest where more computationally intensive methods would be needed to assess the viability of these models \cite{horiuchi16, bozek16, vmbw12, viel05, ssmm12, dekker22, vdle18}.

In Fig.~\ref{fig: contour plot suhalos}, we see that the fraction of sterile neutrino dark matter, $f$, increases as the total mixing angle and initial lepton number increases. The discussion of Fig.~\ref{fig:rowplotlep} and Fig.~\ref{fig:rowplotmix} explain these results. Increasing the mixing angle boosts both resonant and non-resonant production, while increasing the initial lepton number boosts resonant production. 

The number of subhalos generally decrease as $f$ increases. When there are no sterile neutrinos, \mbox{$f=0$}, there are roughly 160 subhalos (as expected from Ref.~\cite{sch15}), which may overproduce these structures. 
We hold the total dark matter density fixed, so as $f$ increases, cold dark matter is replaced by warmer sterile neutrino dark matter. The result is the formation of fewer subhalos. Non-resonant production produces warmer spectra than resonant, so we can see that the number of subhalos decreases more rapidly with increasing mixing angle than with increasing initial lepton number.

We introduce the astrophysical mixing angle, 
\begin{equation}
    \label{eq:astro mix angle}
    \sin^22\theta_{\mathrm{astro}} = f \times \sin^22\theta_{\mathrm{tot}},
\end{equation}
to connect these models to the observed X-ray line. This is the mixing angle that is astrophysically-inferred from the flux of the 3.55 keV line, assuming that 100\% of the dark matter is sterile neutrino dark matter. 
The sterile neutrino decay rate to produce the X-ray line is proportional to $\sin^22\theta_{\mathrm{tot}}$. However, in these mixed sterile neutrino dark matter models, there would be fewer decaying sterile neutrinos in the telescope field of view than assumed. As a result, the inferred decay rate from the X-ray line flux would be less than the sterile neutrino decay rate, and $\sin^22\theta_{\mathrm{astro}} \le \sin^22\theta_{\mathrm{tot}}$. 

Fig.~\ref{fig: scatter plot mixang} depicts the same models as shown in Fig.~\ref{fig: contour plot suhalos}, but with respect to $\sin^22\theta_{\mathrm{astro}}$, $f$, and the number of subhalos, all of which can be probed by astrophysical observations.  The colored dots and cross marks represent models sampled from a linear, rectangular grid in the parameter space shown in Fig.~\ref{fig: contour plot suhalos}.

Astrophysical constraints can be assessed in Fig.~\ref{fig: scatter plot mixang}. Viable models do not overproduce sterile neutrino dark matter, $f \leq 1$, and the lower boundary of models represents the minimum sterile neutrino dark matter produced purely non-resonantly with zero initial lepton number. The color scheme outlines the viable models that produce 60 or more subhalos. Finally, the astrophysically inferred mixing angle is either measured from detections of the X-ray line, or as upper limits from non-detections of the line.

The results of this three-to-one model with equal mixing angles and equal initial lepton numbers show that models with nearly 100\% sterile neutrino dark matter and an estimated 60 subhalos can be created with \mbox{$\sin^2 2 \theta_{\rm astro} \approx 7 \times 10^{-10}$}. This value is consistent with some detections of the X-ray line, but not others (see, {\it e.g.}, Fig.~7 in Ref.~\cite{abazajian2017} and references therein). Furthermore, we see in Fig.~\ref{fig: scatter plot mixang} that these viable mixed sterile neutrino dark matter models are consistent with $\sin^2 2 \theta_{\rm astro} \lesssim 10^{-9}$, which is consistent with all detections of the X-ray line.

\begin{figure}[h!]
	\centering
	\includegraphics[width=9cm]{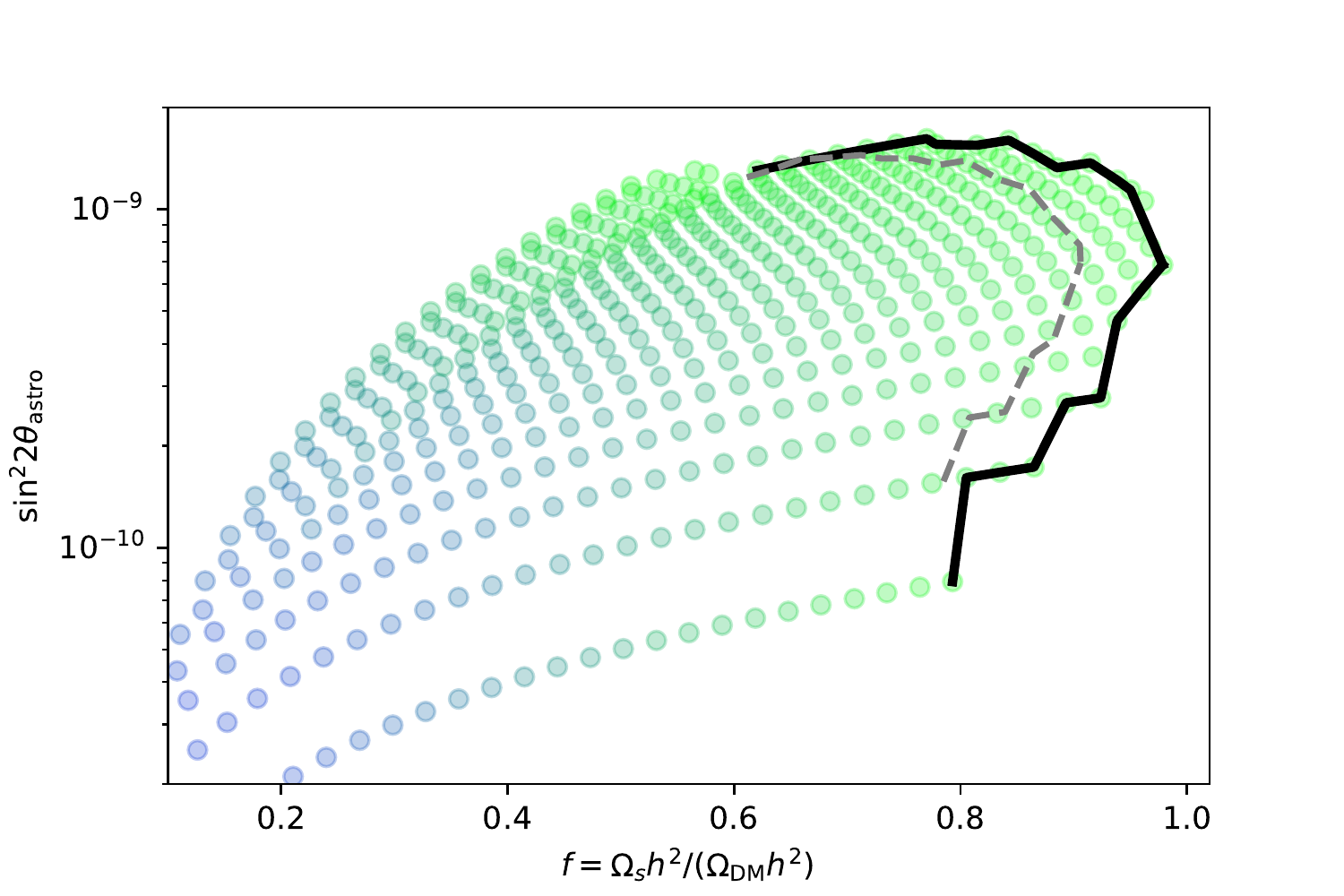}
	\caption{The dots are a reproduction of the viable models from Fig.~\ref{fig: scatter plot mixang} and the solid curve estimates the upper bounds of these three-to-one models with equal mixing angles and equal initial lepton numbers. The dashed curve superimposed on this image is the approximate upper bounds for three distinct three-to-one mixing schema where one flavor has ten times the mixing angle of the other two, while having the same initial lepton number. All three schema have approximately the same upper bounds, so to ease confusion, this just shows the bounds where the tau-to-sterile channel has ten times the mixing angle of the others.}
	\label{fig:new models}
\end{figure}

We can ask how these results change with different schema of mixing angles and/or initial lepton numbers. For each model, we can create models akin to Fig.~\ref{fig: scatter plot mixang} to analyze different models. Fig.~\ref{fig:new models} reproduces the viable models from Fig.~\ref{fig: scatter plot mixang} with a solid line roughly tracing the bounds of viable models in the $\sin^2 2 \theta_{\rm astro}$-$f$ plane. From this outline, we can see the aforementioned results of 100\% sterile neutrino dark matter with \mbox{$\sin^2 2 \theta_{\rm astro} \approx 7 \times 10^{-10}$} and a maximum \mbox{$\sin^2 2 \theta_{\rm astro} \lesssim 10^{-9}$}.

The dashed curve in Fig.~\ref{fig:new models} is the approximate bounds of the viable models for 3 distinct three-to-one mixing schema where one flavor has ten times the mixing of the other two with the same initial lepton numbers. These curves have overall similar shapes, but the details are interesting. First, we notice that the maximum fraction of sterile neutrino dark matter in viable models is smaller in unequal mixing schema. This noticeable, but small effect is likely due to the fact that more resonant production in any given channel creates slightly warmer spectra as the resonance sweeps to higher $\epsilon$ values. Second, the maximum allowed astrophysically-inferred mixing angle is roughly unchanged between the mixing schema. This is likely dominated by the warmer non-resonant production whose overall production combined from all three channels is proportional to $\sin^2 2 \theta_{\rm tot}$.

We conclude with some caveats to these results which lead us to future work. In this work, we focused on the properties of our three-to-one mixed sterile neutrino dark matter model, where the lepton number-driven transformation of all three active neutrino species into sterile neutrinos is complemented by cold dark matter. In particular, we explored the properties of the spectra produced in these models and their connection with astrophysical constraints. 

In our exploration, we used the scattering rates from Ref.~\cite{afp} instead of those used in more sophisticated calculations, for example as in Ref.~\cite{vcah16}. While updated scattering rates will change the specifics of the conclusions, the overall properties and discussion of these mixed sterile neutrino dark matter models will remain unchanged. For example, with the higher scattering rate in Ref.~\cite{vcah16}, we expect the boundary curve in Fig.~\ref{fig:new models} to have a similar shape, but the maximum $\sin^2 2 \theta_{\rm astro}$ will be lower and the maximum fraction of sterile neutrino dark matter will be lower as well. This is because the higher scattering rate will boost non-resonant production and suppress resonant production.

Another question involves mixing between the three active neutrinos. In principle, a $4 \times 4$-mixing matrix would describe the neutrino sector of these mixed sterile neutrino dark matter models \cite{gdp19}. However, because the mixing angles considered are so small, there is negligible loss in independently approaching transformation in each active-sterile channel. The $3 \times 3$ mixing between the active neutrinos (and the anti-neutrinos) may present non-trivial quantum kinetic evolution, especially when the lepton numbers in the three flavors are different \cite{abb02, abfw05}. While it is possible that resonances can efficiently redistribute and equilibrate the lepton numbers amongst the three flavors, the high scattering rates at \mbox{$T \sim 300~{\rm MeV}$} may suppress this process. Nevertheless, if we look at Fig.~\ref{fig: lep evo 3 to 1}, we see that in the three-to-one model with equal mixing angles and lepton numbers, the three lepton numbers have similar evolutions, so we feel that this may have little effect on these models. However, if we investigate mixing schema with different mixing angles for the three channels, this may cause the lepton number evolution of the different flavors to diverge, and this effect may become more pronounced.

Structure formation is a valuable approach to constrain dark matter production models. In this work we used an approach to estimate the number of subhalos produced and constrain these models in connection with the number of observed Milky Way satellites (see, {\it e.g.}, Refs.~\cite{sch15, ch17}). This approach doesn’t appear to be grossly out of line with other structure-formation based approaches to constraining dark matter spectra \cite{zelko22, sch16}. While we do not expect any drastic changes to the conclusions drawn in this work, other structure formation-based approaches should be incorporated when discussing models at the edge of viability as discussed in Fig.~\ref{fig:new models}.

\acknowledgements
The authors would like to acknowledge support from NSF grant PHY-2111591 and from the College of Arts and Sciences at the University of San Diego.

\bibliography{refs}

\end{document}